# Image-based ground distance detection for crop-residue-covered soil

**Baochao WANG [a,*], Xingyu ZHANG[a], Qingtao ZONG[a], Alim PULATOV[b], Shuqi SHANG[a], Dongwei WANG[c]**

[a] College of Mechanical and Electrical Engineering, Qingdao Agricultural University, Qingdao 266109, China

[b] National Research University "Tashkent Institute of Irrigation and Agricultural Mechanization Engineers", Tashkent, 100000, Uzbekistan

[c] Yellow River Delta Intelligent Agricultural Machinery Equipment Industry Academy

## Abstract

Conservation agriculture features a soil surface covered with crop residues, which brings benefits of improving soil health and saving water. However, one significant challenge in conservation agriculture lies in precisely controlling the seeding depth on the soil covered with crop residues. This is constrained by the lack of ground distance information, since current distance measurement techniques, like laser, ultrasonic, or mechanical displacement sensors, are incapable of differentiating whether the distance information comes from the residue or the soil. This paper presents an image-based method to get the ground distance information for the crop-residues-covered soil. This method is performed with 3D camera and RGB camera, obtaining depth image and color image at the same time. The color image is used to distinguish the different areas of residues and soil and finally generates a mask image. The mask image is applied to the depth image so that only the soil area depth information can be used to calculate the ground distance, and residue areas can be recognized and excluded from ground distance detection. Experimentation shows that this distance measurement method is feasible for real-time implementation, and the measurement error is within ±3mm. It can be applied in conservation agriculture machinery for precision depth seeding, as well as other depth-control-demanding applications like transplant or tillage.

## 1. Introduction

Conservation agriculture technique has the characteristics of minimal soil disturbance and crop residues remaining on the soil surface, and the soil surface covered with crop residues has multiple advantages such as water saving, soil conservation and ecological protection (Zhou et al.,2020; Deng et al.,2022; Wen et al.,2023; Jahangirpour et al.,2023; Zhang et al.,2024). However, no-till and stubble mulch can have an impact on some operations of the agriculture machine, such as crop residues affecting ground distance measurement and associated depth control, thus traditional planters cannot achieve precision depth sowing.

For traditional agricultural machinery, there are two primary adjustable depth setting strategies. The first is a pressure-based control strategy. In this approach, constant depth setting could be accomplished by relying on the constant downward force generated by a hydraulic or electric actuator. For instance, models like John Deere 1725C and LEMKEN Solitair 9 follow this approach (John Deere,2024; LEMKEN,2024). However, this kind of depth control is indirect and can be influenced by various factors such as soil type, soil moisture, residues and other parameters (Jing et al.,2020; Virk et al.,2021). The second depth setting strategy for traditional agricultural machinery relies on sensor measurement. Sensors such as ultrasonic, laser, or other types can be used. This approach offers a more direct measurement of the depth compared to the pressure-based control strategy.

However, the above downward-force-based and sensor-based approaches cannot work well in conservation agriculture where the soil surface is covered with crop residues. For the downward-force-based depth adjusting techniques, it is only suitable for pre-cultivated seedbeds, and in the case of soil surfaces with residues, the force generated by pressing on the residues cannot reflect the relationship with depth. The sensor-based depth adjusting techniques also encounter a similar problem. Laser and ultrasonic sensors cannot distinguish the distance measured from the soil or residue in conservation agriculture. As a result, they are not suitable for conservation agriculture either.

More specifically, existing depth determination strategy based on sensor measurement mainly includes two types: direct ground distance measurement and indirect ground distance measurement.

The direct ground distance measurement method uses laser, ultrasonic and optical sensors to measure the distance directly (Søgaard 1998; Kiani et al.,2010; Wen et al.,2014; Sharipov et al.,2018; Mapoka et al., 2019; Machleb et al.,2020; Li et al.,2023; Ye et al.,2023). LEE et al. designed a detection unit for a tillage depth control by installing optical and ultrasonic sensors on a tractor to measure ground distance (LEE et al.,1998). VAN DER LINDEN et al. developed a depth sensor based on infrared laser triangulation to measure the distance between the machinery frame and the ground in real-time, validated through experiments to be unaffected by soil structure and sunlight (VAN DER LINDEN et al.,2008). Suomi and Oksanen measured the relative distance between the frame and the soil surface by installing ultrasonic sensors on the frame (Suomi and Oksanen,2015). NIELSEN et al. developed a measurement system using ultrasonic sensors to measure the distance between the frame and the ground (NIELSEN et al.,2017). Mohammadi et al. used infrared sensors and ultrasonic sensors to measure the distance change of soil surface at different driving speeds, the results showed that the infrared sensors can quickly measure changes in soil surface distance, while ultrasonic sensors are inaccurate (Mohammadi et al.,2023).

NIELSEN et al. used a position sensor combined with an ultrasonic sensor to measure the furrowing depth of the plow blade (NIELSEN et al.,2018).

The indirect ground distance measurement is carried out using mechanical geometric relationships and mechanical displacement sensors mounted on machines (JIA et al.,2016). JIA et al. calculated the tillage depth by measuring the angle between the implement and the swing arm (JIA et al.,2016). MOUAZEN et al. developed a pendulum-type sensor that calculates the relative distance to the ground by measuring the distance between the frame and the wheel axle (MOUAZEN et al., 2004).

However, the above study cannot work well in conservation agriculture where the soil surface is covered with crop residues. Because the existing measurement methods are unable to distinguish the influence of crop residue from that of the soil. (LEE et al.,1996; MOUAZEN et al.,2004; Saeys et al.,2004). Saeys et al. showed that on the ground where there is a corn crop, the ultrasonic sensor's signal is reflected from the corn crop rather than from the ground (Saeys et al.,2004).

Thus, facing the existing problems of detecting ground distance in conservation agriculture, this paper has proposed a solution with the ability to distinguish the distance information from the soil or crop residues. This method is performed with 3D camera and RGB camera, obtaining depth image and color image at the same time. The distance information of a large area is obtained by the 3D camera. The color image is used to distinguish the different areas of residues and soil and finally generates a mask. The mask is applied to the depth image so that only the soil area depth information can be used to calculate the ground distance, and residue areas can be recognized and excluded from ground distance detection.

## 2. Materials and Methods

### 2.1 General principle and method

The schematic diagram of the proposed method is shown in Fig. 1. The ground distance is obtained from images acquired by 3D cameras and RGB cameras. The 3D camera image acquires distance image including distance information, and the RGB camera acquires the color image. The two images are processed and aligned. Then, the color image is used to distinguish the crop-residue and the ground for generating a binary mask. The mask is then applied to the depth image, and finally the ground distance information after removing the crop residue influence can be calculated.

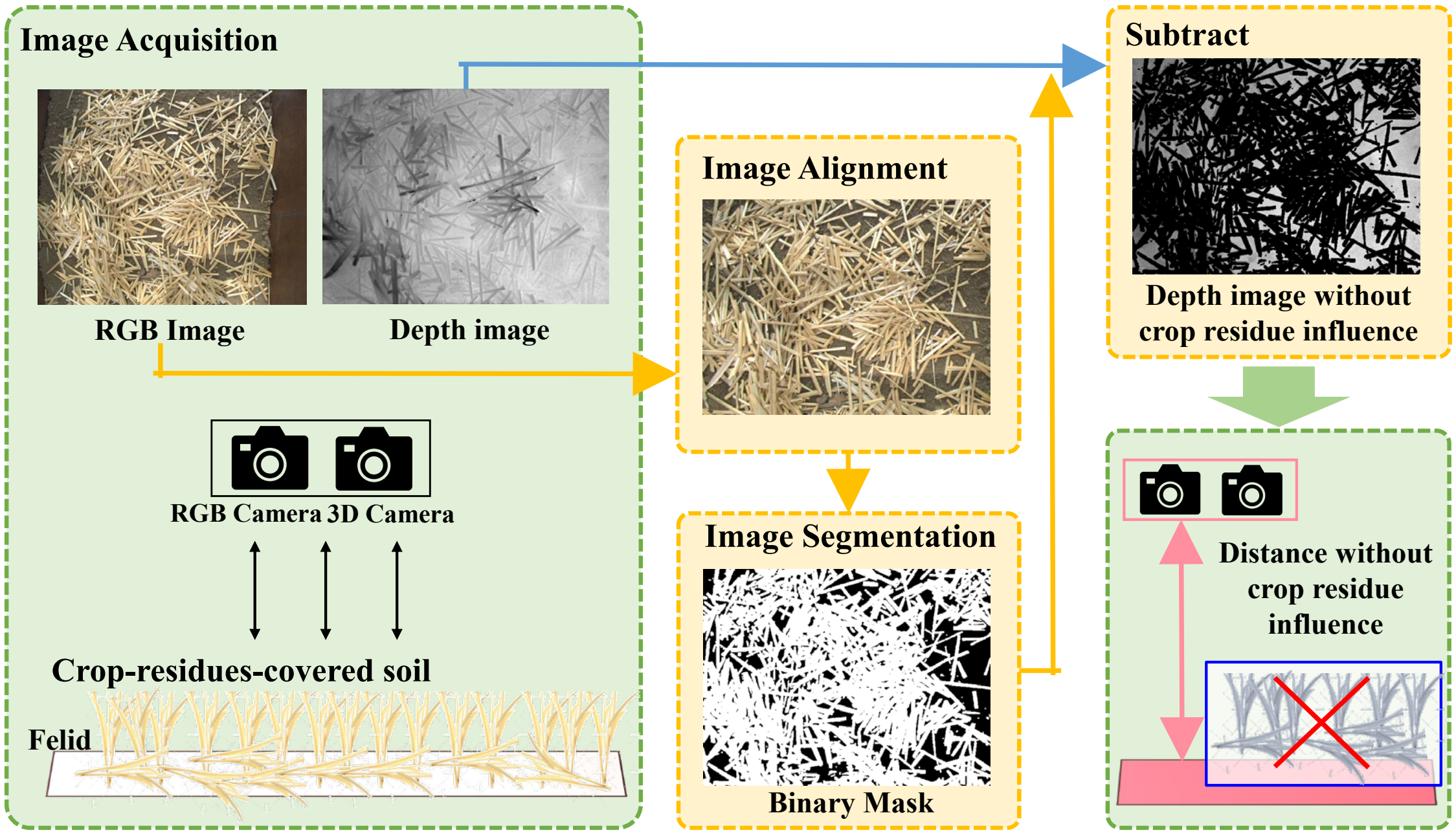

Fig. 1 Ground distance detection schematic diagram

### 2.2 Hardware configuration

The hardware includes 3D camera, RGB camera, and Industrial PC (IPC). The 3D camera is Basler Blaze-101, the RGB camera used is Basler Aca1300-75gc. The two cameras are combined together with a plastic frame, as shown in Fig. 2. The IPC is Beckhoff C6030-0060. The key parameters of the hardware are shown in Table 1.

Table1 Key parameters of hardware components

| | Name | Value |
|---|---|---|
| 3D Camera Basler Blaze-101 | Depth information acquisition principle | Time of flight (TOF) |
| | Working range | 0.3~10m |
| | Interface | GigE |
| | Resolution | 640×480 |
| | Accuracy (typical) | ±5mm(0.5~5.5m) |
| | Field of View | 67°×51° |
| | Illumination | VCSEL, 940nm |
| | Frame Rate | 20fps(default)<br>30fps(fast mode) |
| RGB Camera | Interface | GigE |
| | Resolution | 1280×1024 |
| | Sensor Format | 1/2" |
| | Effective Sensor Diagonal | 7.9 mm |
| | Frame Rate | 88fps(fast mode)<br>81fps(normal mode) |
| IPC Beckhoff C6030 | CPU | Intel i7-7700,<br>3.6GHz, 4 cores |
| | Ethernet | 4×100/1000BASE-T |
| | Operating system | Windows 10 IoT<br>Enterprise |
| Camera integration | Dimension (W×H×L) | 140×81×63 mm |

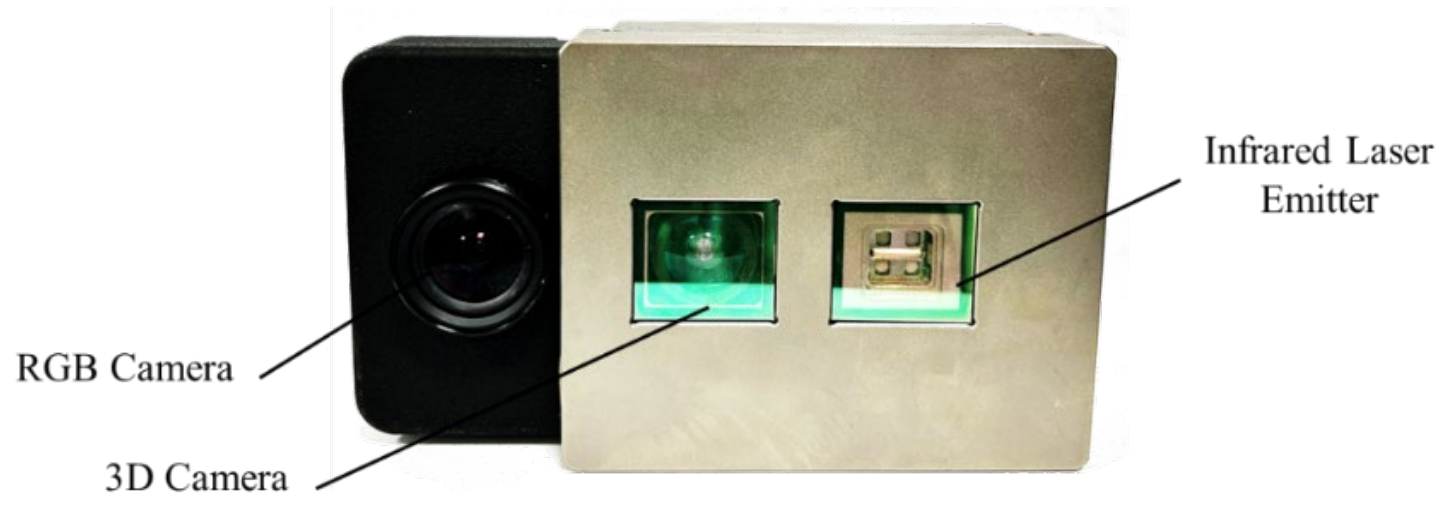

Fig. 2 Camera combination

### 2.3 Depth image acquisition and processing

The working principle of a 3D Time of flight (TOF) camera is to calculate the distance between the measured object and the camera by measuring the time difference from the emission of a light pulse to its reflection back to the camera. A 3D camera is capable of providing two types of images: an optical image that is derived from the intensity of light, and a distance image which is generated based on the distance calculated from the flight time of light.

The first step is to process the distance image acquired by 3D camera, in order to get depth image only including vertical ground distance information. In the mono color image obtained by the 3D camera, each pixel offers a 16-bit gray value that spans from 0 to 65535, representing the original distance (encompassing the x, y, and z directions) between the camera and the target. For our application, it is required to use the vertical distance (the distance only includes $z$ direction). Thus, a calculation is required to get depth image from original image, as in eq (1).

$$z = g \times \mu \times \sqrt{\frac{1}{\left(\frac{u-cx}{f}\right)^2 + \left(\frac{v-cy}{f}\right)^2 + 1}} - z_{offset} \tag{1}$$

where $g$ is the pixel gray value; $\mu$ is the scale factor between gray value and real distance, which is 0.0229 for the used camera; $u$ is the abscissa size of depth image, ranging from 1 to 640 for the used camera; $v$ is the ordinate size of depth image, ranging from 1 to 480 for the used camera; $f$ is the focal length of 3D camera, which is 509.935 pixels for the used camera; $cx$ is the abscissa of the optical center, which is 313.05 for the used camera; $cy$ is the ordinate of the optical center, which is 239.60 for the used camera; $z_{offset}$ is the z-axis offset for compensating the displacement between the internal sensor plane and the camera mounting surface, which is 23.97mm for the used camera.

### 2.4 Image alignment

Because the resolution of 3D camera and RGB camera is different, it is necessary to align RGB image and depth image before image processing. The coordinate systems of different images are converted to the same by Formula (2~5), so that the same pixel positions of RGB image and the depth image are in correspondence.

$$M = \begin{bmatrix} R & T \end{bmatrix} \tag{2}$$

$$W = K_d^{-1} \bullet M \bullet K_c \tag{3}$$

$$u_{(x,y)} = x \bullet w_{11} + y \bullet w_{12} + w_{13} + \frac{w_{14}}{z_{(x,y)}} \tag{4}$$

$$v_{(x,y)} = x \bullet w_{21} + y \bullet w_{22} + w_{23} + \frac{w_{24}}{z_{(x,y)}} \tag{5}$$

where $M$ is the external parameter matrix; $R$ is the rotation matrix; $T$ is a translation matrix; $K_d$ is the internal parameter matrix of 3D camera; $K_c$ is the internal parameter matrix of RGB camera; $u$ is the abscissa of the optical center in RGB image; $v$ is the ordinate of the optical center in RGB image; $x$ is the abscissa of the optical center in depth image; $y$ is the ordinate of the optical center in depth image. $M$, $K_d$ and $K_c$ are obtained by the camera calibration method proposed by Zhang(Zhang,2000). The images before and after alignment are shown in Fig. 3.

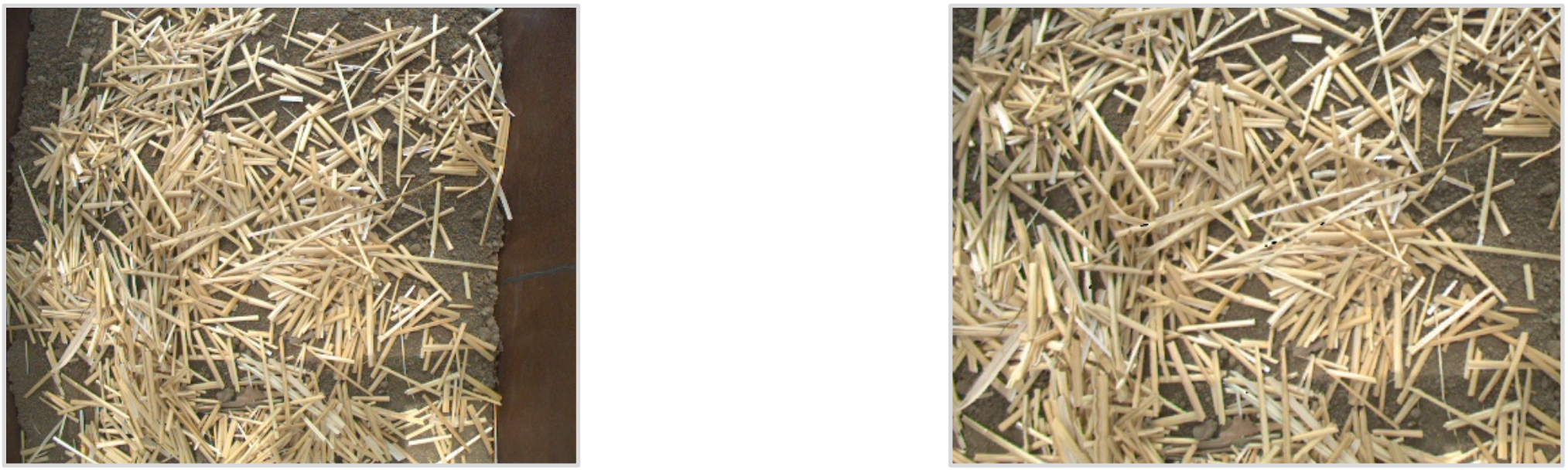

(a) Original RGB image (b) Aligned RGB image

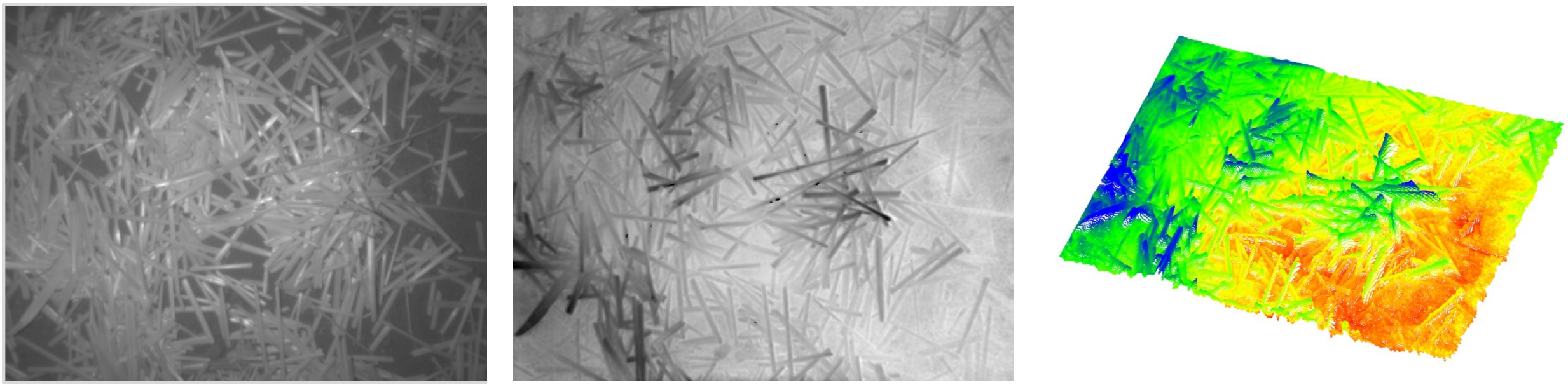

(c) Depth image (optical image by light density) (d) Depth image (gray value image by distance) （e）Depth image represented in the form of point cloud

Fig. 3 RGB image before and after alignment

### 2.4 Image segmentation and combination

After alignment, the image segmentation is performed to remove the image areas influenced by crop residues, so as to generate a binary mask to remove crop residue influence. The process involves several steps such as color segmentation, binarization, dilation operation, subtraction operation. After these steps are completed, the final output is the depth image without the influence of crop residues covering.

Firstly, through implementing different color thresholds in the RGB image, ground and crop residues can be effectively separated. Secondly, the RGB image with the crop residues removed is binarized to generate a mask. Subsequently, the binarized mask is applied to the depth image for subtracting the area influenced by the crop residues. Finally, the ground distance value is computed based on the processed depth image.

Fig. 4 (a) gives original RGB image. Fig. 4 (b) is the depth image after removing the covering crop residues. The black part in Fig. 4 (b) is the area of covering crop residues, and the gray part is the vertical ground distance information. By comparing with RGB image in Fig. 4 (a), it can be seen that the image segmentation method can realize the separation of the ground and the covering crop residues, and thus effectively getting ground distance information.

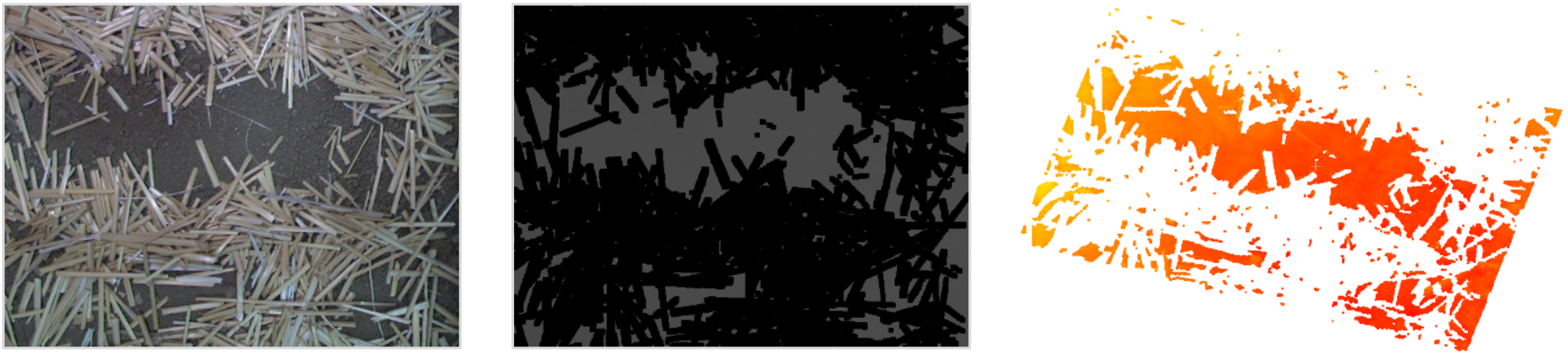

(a) RGB Image

(b) Depth image after removal of residues

(c) Depth image after removal of residues (in form of point cloud)

Fig. 4 Image processing results

## 3. Results

The proposed ground distance detection method is evaluated experimentally by static test, dynamic test and field test.

### 3.1 Static test

To verify the feasibility and to evaluate accuracy of the proposed ground distance detection method, an experimental rig for static test was firstly set up, as shown in Fig. 5. In the experiment rig, the complex ground environment for conservation agriculture is emulated with wheat stubble and mulching. The static tests were performed with diverse cases in which the residues coverage levels were different.

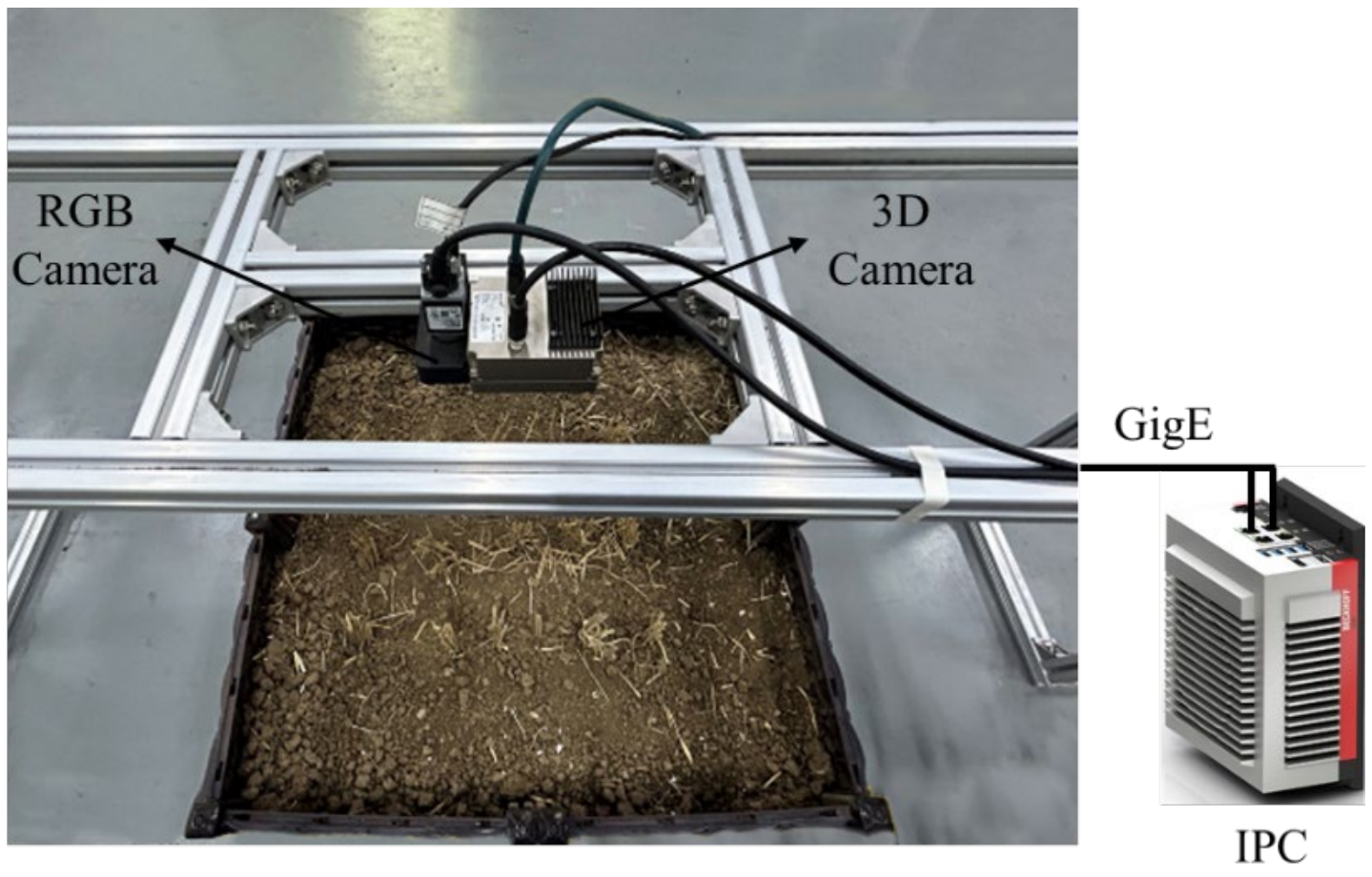

Fig. 5 Ground distance measurement test diagram under static conditions

In the first case (CASE A), the soil is configured only with stubbles. The configuration for CASE A is shown in Fig. 6 (a), where the stubble height is around 60~70mm. Fig. 6 (a) also presents the actual ground distance that was measured with the measuring tape, which is around 518.0 mm. Fig. 6 (b) shows the ground distance detection result by the proposed method after removing the covering, which is 517.2mm. Compared with the actual distance (518.0mm), the error is within 1mm. This result preliminarily demonstrating the feasibility of the ground distance detection method.

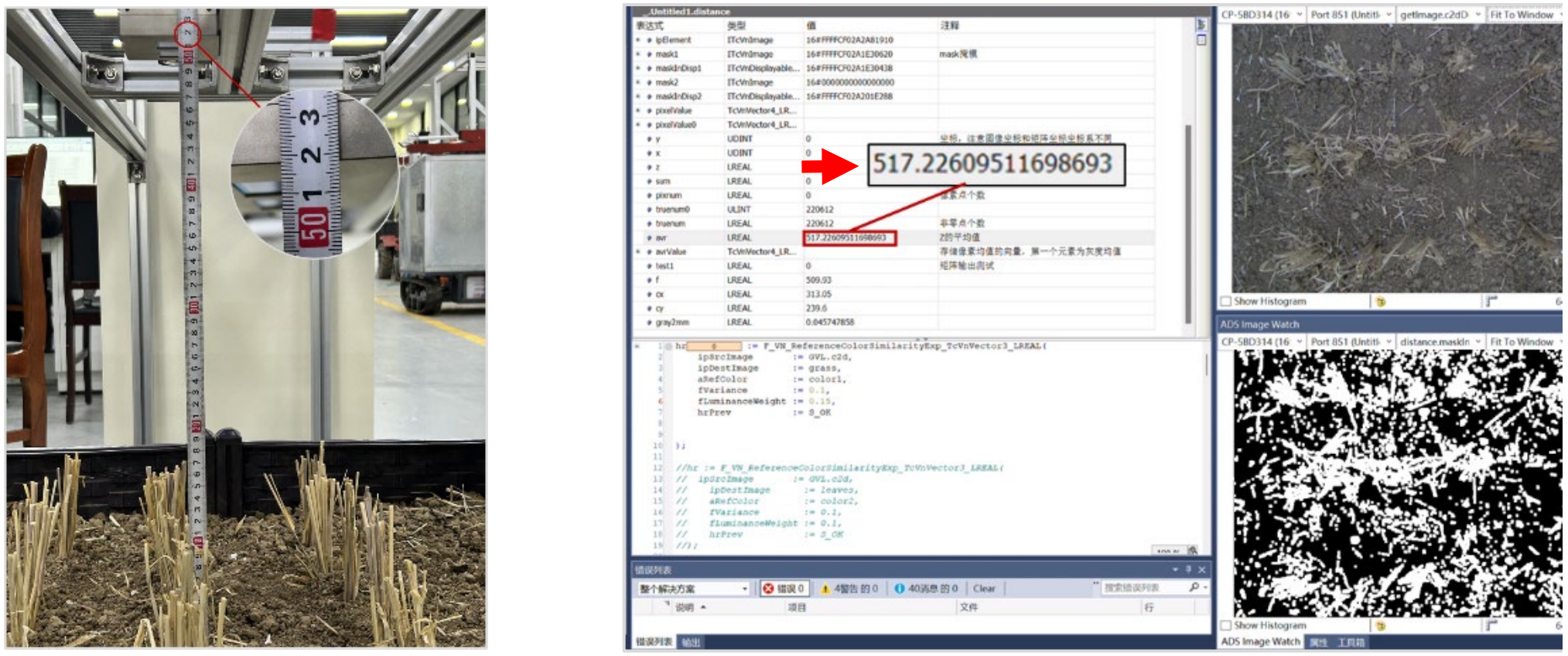

(a) Configuration and actual distance

(b) Ground distance detection results

Fig. 6 Ground distance detection results for CASE A

Based on CASE A, subsequent cases (CASE B~E) are conducted by adding more crop residues without disturbing the soil, so the actual ground distance could remain the same. The key intermediate results for CASE A~E are shown in Fig. 7. In Fig. 7, it can be observed that for different levels of coverage, the proposed method has the ability to distinguish between the residue and the ground. Even the ground area is decreasing with more residue coverage, the method can provide enough data to calculate the ground distance. The drawback lies in the fact that once the ground is entirely covered, the ground distance cannot be calculated effectively. However, in the context of conservation agriculture, the shape and physical characteristics of the mulching are such that it is hardly possible for the surface to be fully covered. Even if some local areas are fully covered in extreme, the invalid frames can be ignored. With the motion of the whole machine, the soil surface distance can also be detected. Consequently, the method proves to be effective for distance calculation.

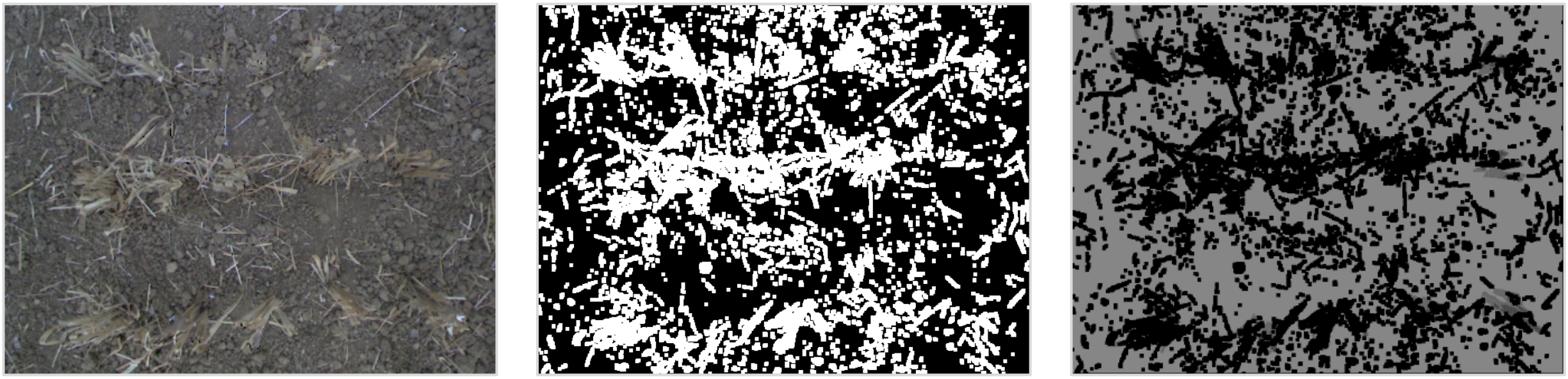

(a) CASE A

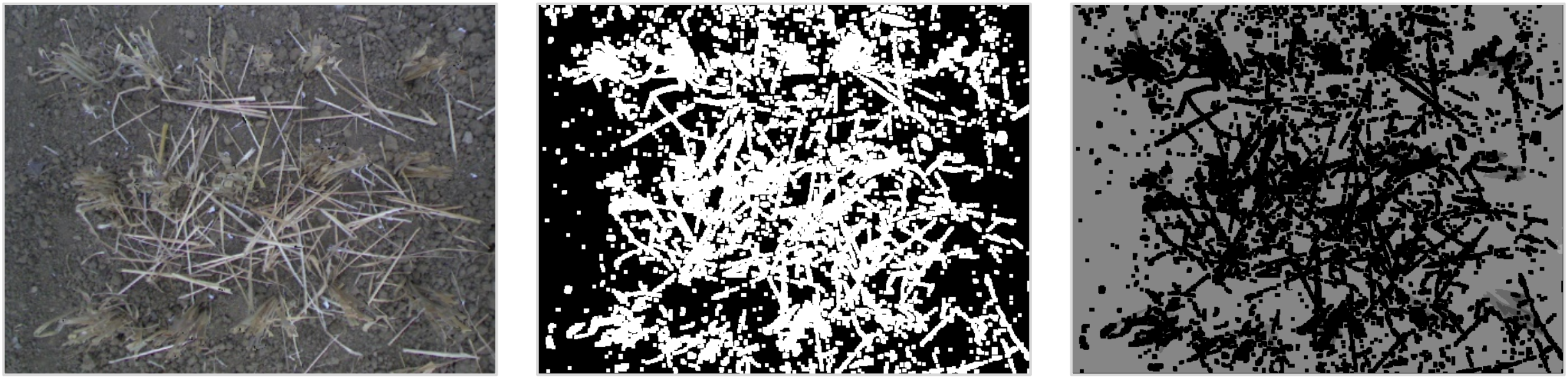

(b) CASE B

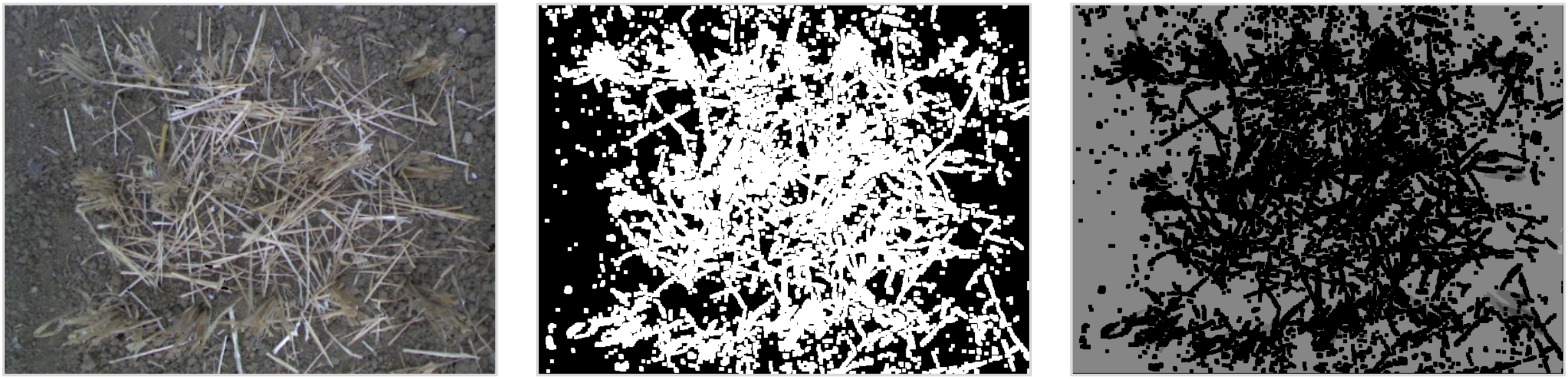

(c) CASE C

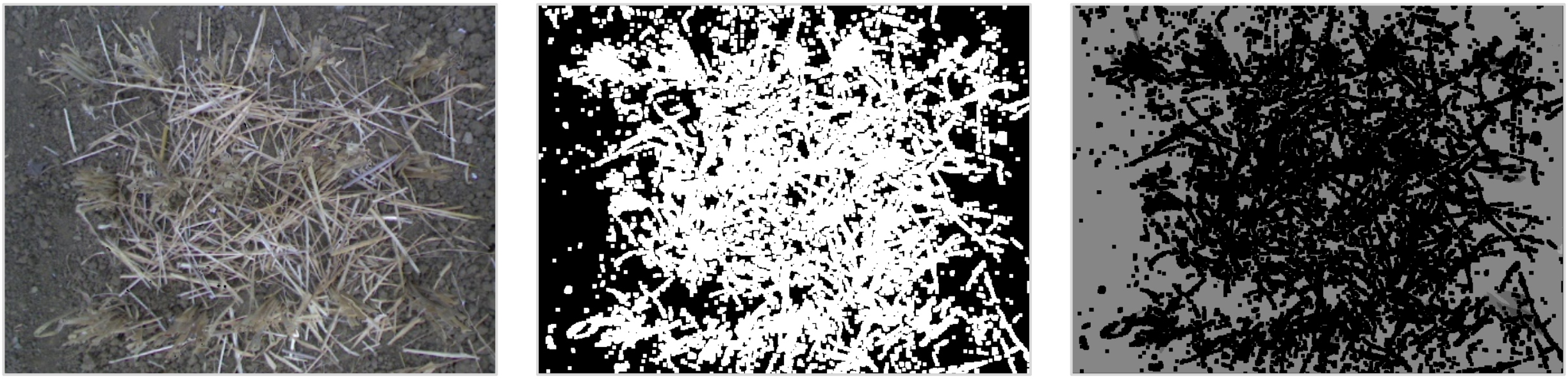

(d) CASE D

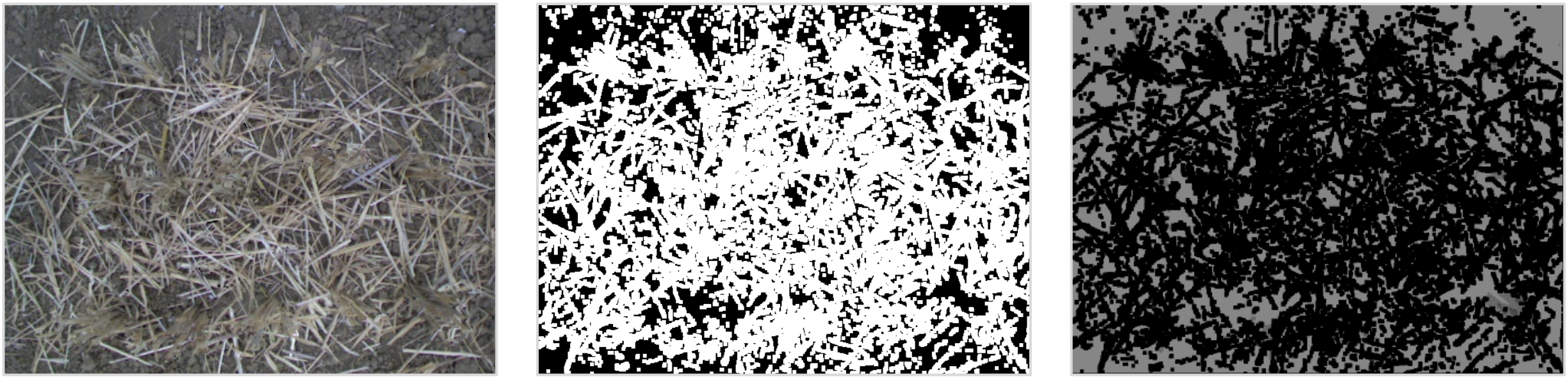

(e) CASE E

Fig. 7 Intermediate key results for different cases (From left to right: RGB image, Binary mask, Depth image after removing the residue influence)

The final results for CASE A~E are summarized in Table 2. From Table 2, it can be observed the detected ground distance maximum error is 2.2mm. In terms of statistics, the average error is -1.66 mm, the standard deviation of the errors in the detection results is 0.523 mm, and the 95% confidence interval of the errors is [-2.319, -1.001]. Given the diameter of a single wheat straw is often above 5mm, after stacking together, the total residue thickness is usually more than 20mm, so the error around 2mm could indicate feasibility and performance of the proposed method of removing crop residue.

On the other hand, the possible cause of the slight variation error in different cases may also come from the fact that the soil surface is not perfectly flat, after the crop residue covers different area in different cases, the rest of the soil area could be different and could cause variations.

Table 2 Ground distance detection under different amount of residue

| | Distance detection results mm | Actual distance mm | Error mm |
|---|---|---|---|
| CASE A | 517.2 | | -0.8 |
| CASE B | 516.4 | | -1.6 |
| CASE C | 516.2 | 518 | -1.8 |
| CASE D | 516.1 | | -1.9 |
| CASE E | 515.8 | | -2.2 |

### 3.2 Dynamic test

To verify the performance of the proposed method in real-time, a circular experimental platform was established. The cameras are fixed on the stationary frame and the circular experimental platform was rotated by four motors. On the rotating circular platform, a layered soil structure with 20mm, 40mm, 60mm and 80mm thickness was constructed. The total length of the layered soil structure is 4000mm. The experiment setup is shown in Fig. 8.

During different cases for the test, different amount of wheat straw is spread on the soil layer to simulate the complex ground surface in conservation agriculture. The camera is placed at a position of 455 mm from the bottom of the circular experiment platform, and the ground distance change is detected in real time by the RGB and 3D cameras using the proposed method, and the detected ground distance *vs.* time is shown in following figures.

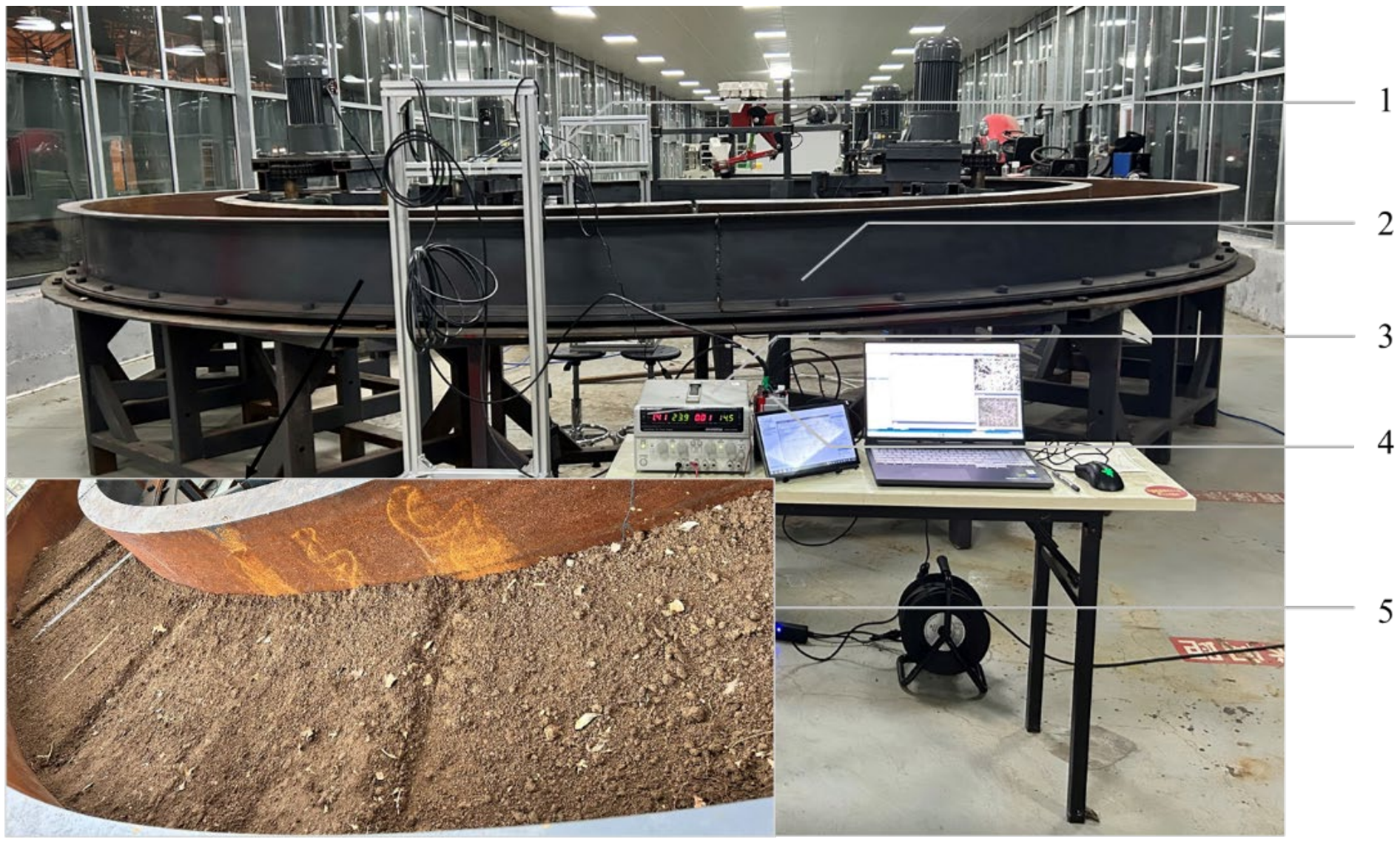

1. 3D Camera and RGB Camera 2. Circular experimental platform 3. DC power supply 4. IPC 5. The layered soil structure

Fig. 8 Experiment rig for of dynamic test

Fig. 9 presents the dynamic testing configuration and results of two cases with different quantities of crop residue. The distribution of crop residues for CASE F and CASE G is shown in Fig. 9 (a) and Fig. 9 (b) respectively. The results are shown in Fig. 9 (c). In Fig. 9 (c), the red curve stands for the original ground distance detected prior to spreading the residue, whereas the black and blue curves denote the detected ground distance after spreading the residues.

From Fig. 9 (c), it is observed that the waveforms are quite similar, with the error being within 2mm, and there is a phase delay among different curves. The delay is mainly caused by the fact that our experiment platform is driven by four induction motors with open loop control, and the speeds cannot be exactly the same in each test. The consistency of the three curves validates the feasibility of the proposed method in real-time.

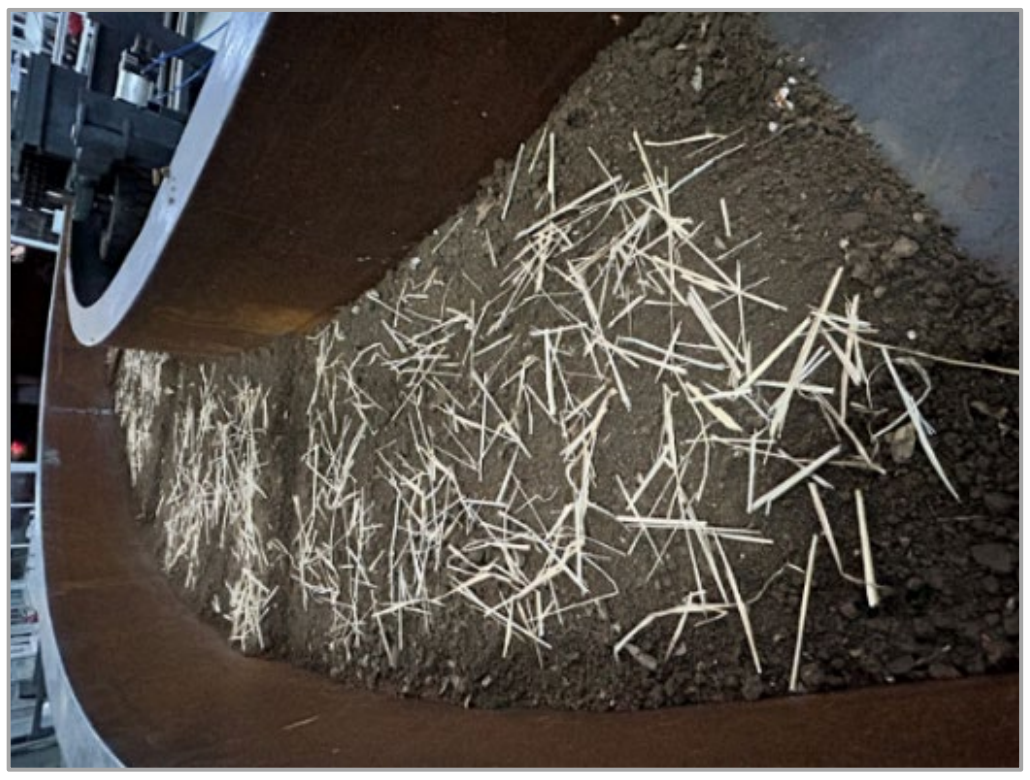

(a) Case F: sparse straw distribution

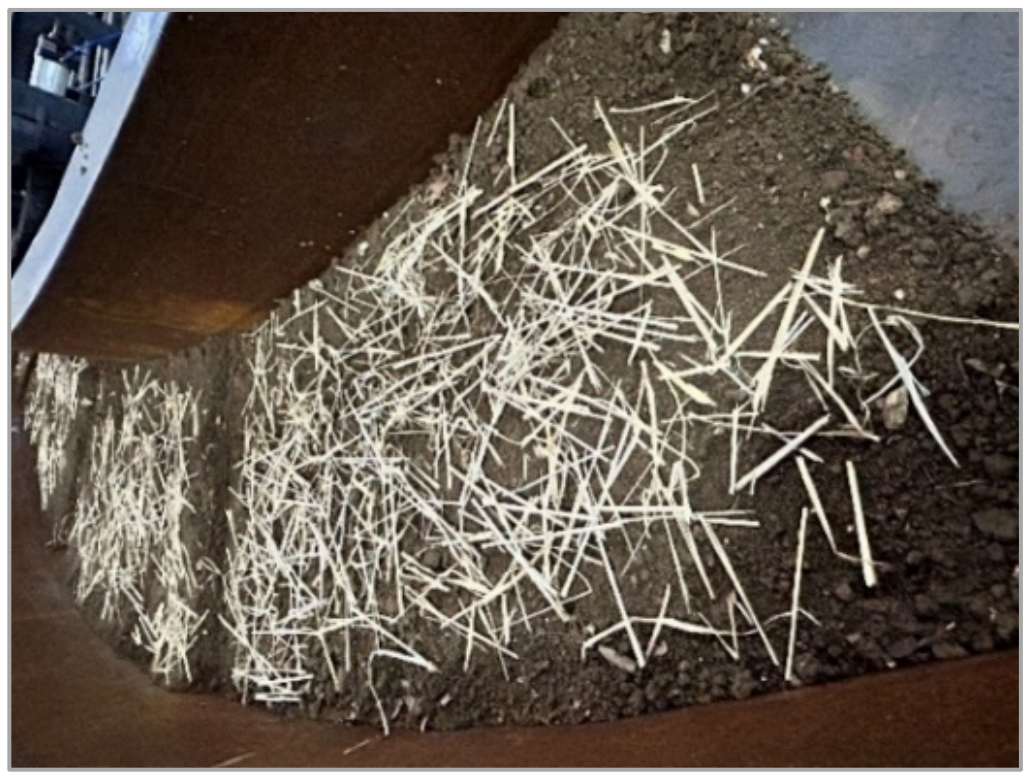

(b) CASE G: thin straw distribution

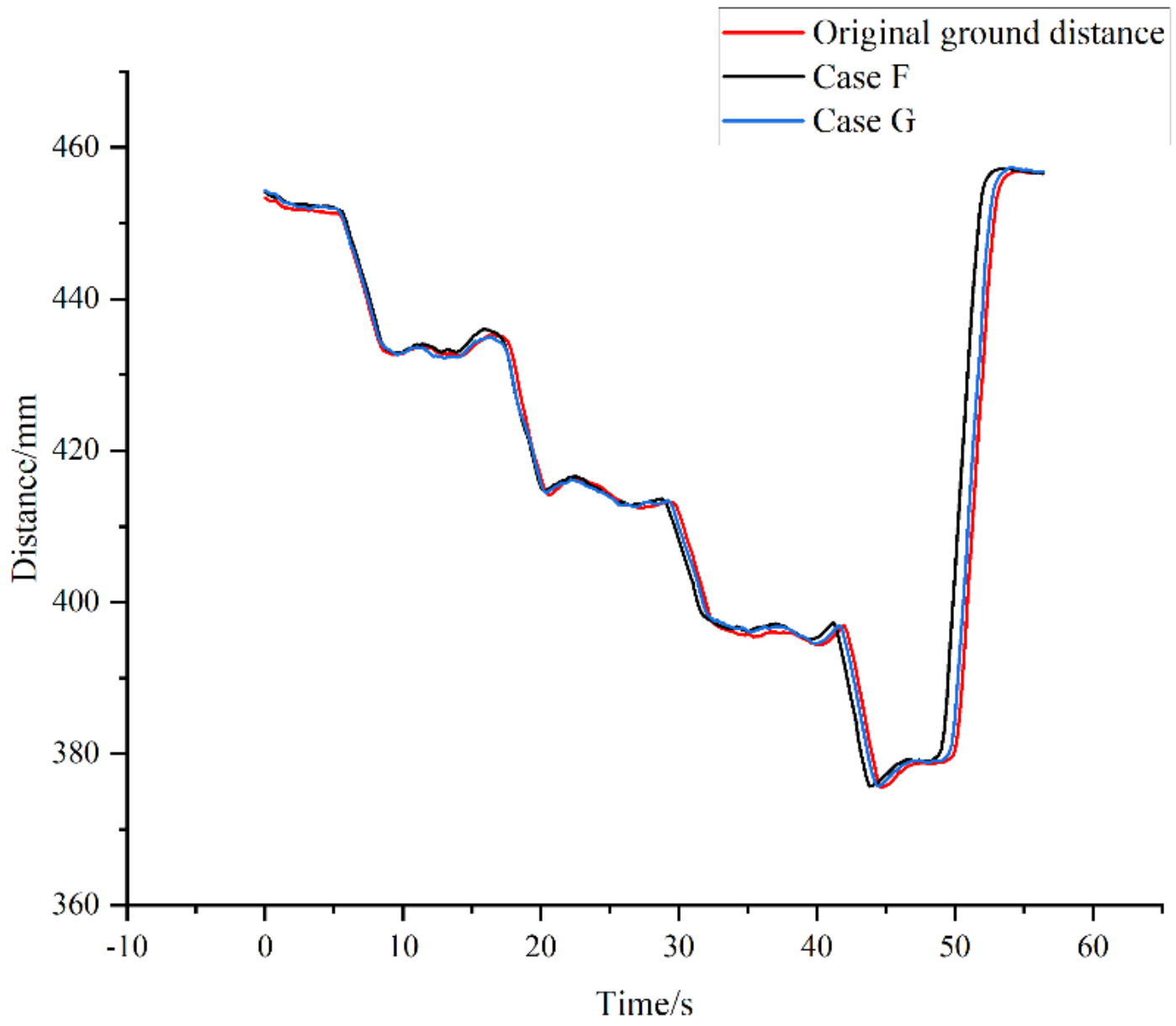

(c) Ground distance detection results

Fig. 9 Ground distance detection for small amount of crop residue coverage

To further explore the performance with thick coverage of residues, all the residues were collected and gradually increased on the 20mm-thick soil layer for test with large amount residue condition, as shown in Fig.11 (a) and Fig.11(b). The detected ground distance is shown in Fig.11 (c).

From Fig. 10 (c), it can be seen that the three curves are in consistent before 7.5s and the error is within 1mm. The error start to increases after 7.5s, and the maximum error is 3mm compared with the case without crop residue. The cause for error is also discussed in following paragraph.

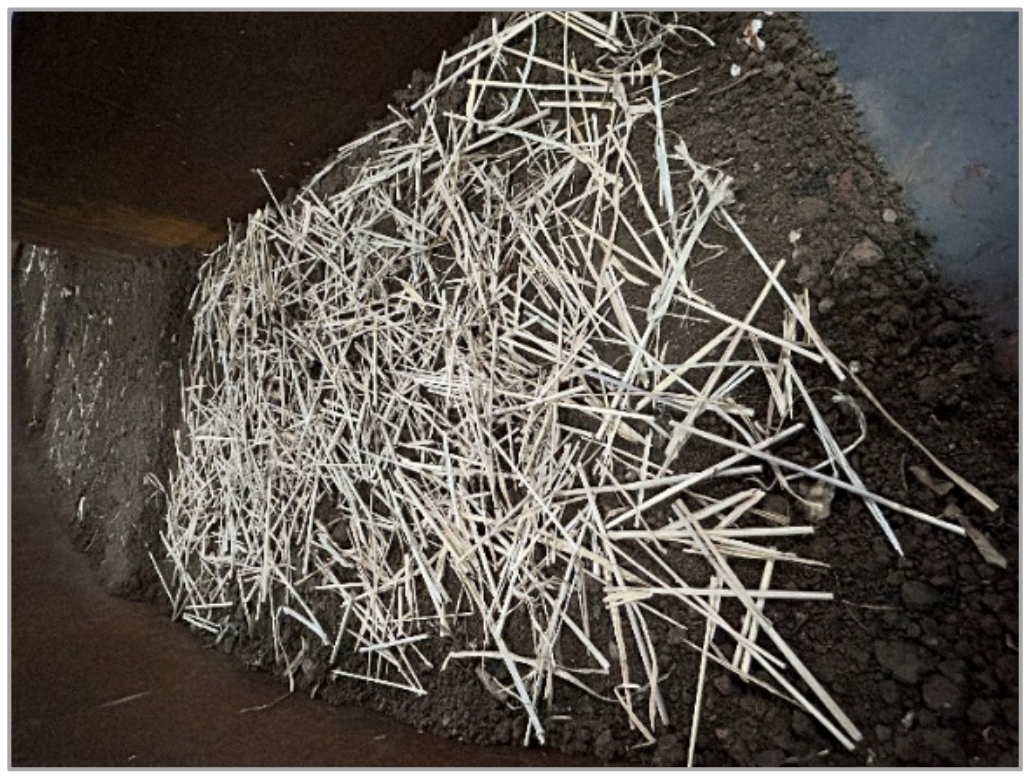

(a) CASE H: thick straws distribution

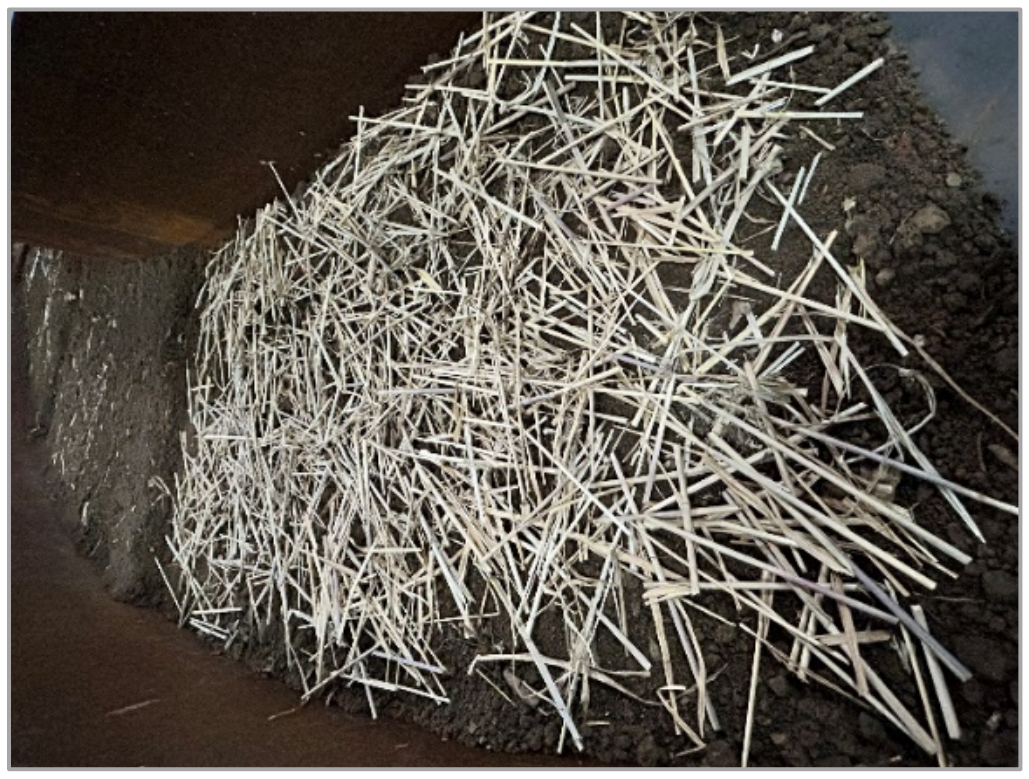

(b) CASE I: much thick straw distribution

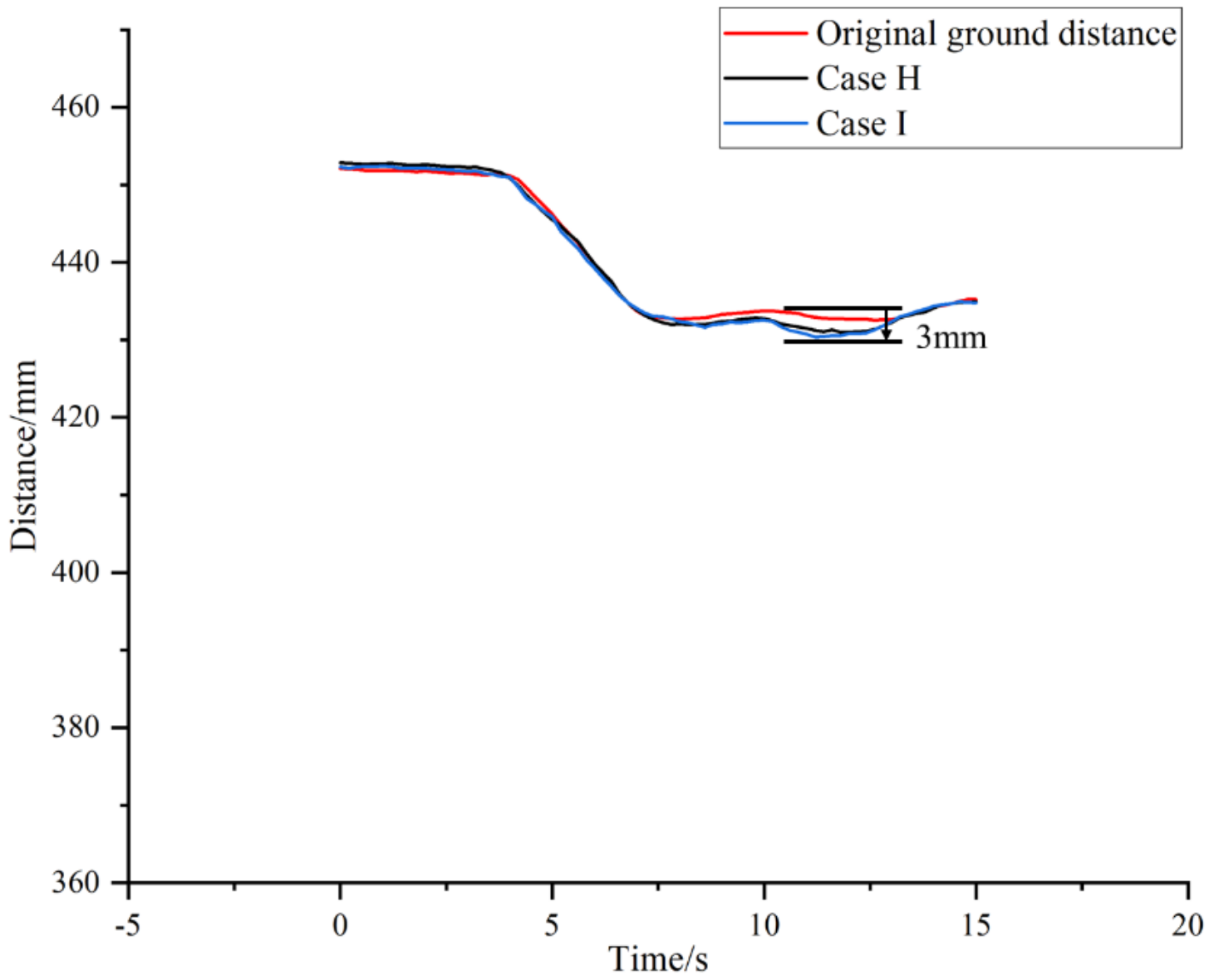

(c) Ground distance detection results

Fig. 10 Ground distance detection for large amount of crop residue coverage

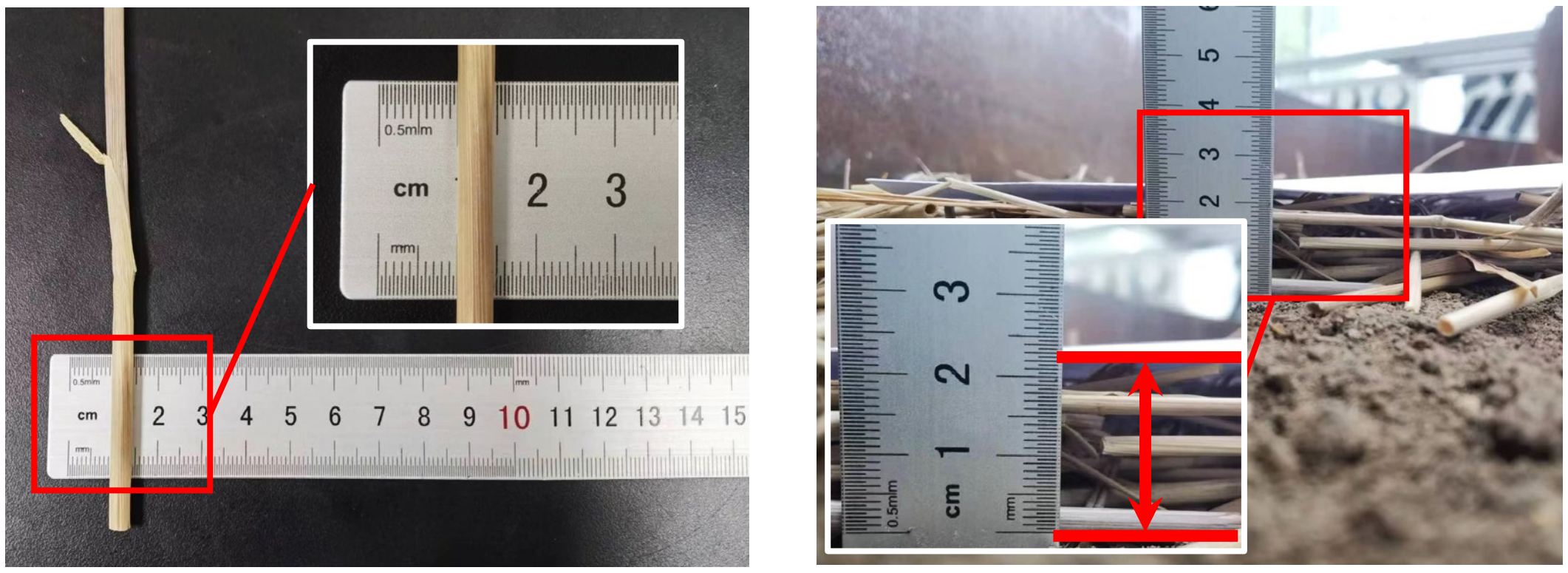

(a) Diameter of a single straw (b) Thickness of residue coverage

Fig. 11 Feasibility proof from measurement of straw and residue coverage

The error is not from the crop residue, since a single straw diameter is more than 3mm and the residue cover thickness is about 22mm, as shown in Fig. 11. As analyzed before, this result may come from the fact that the soil surface is not perfectly flat, after the crop residue covers different area in different cases, the rest of the soil area could be different and could cause variations. The results indicate that the method is feasible and stable for distance detection under different levels of wheat straw coverage.

**3.3 Field test**

The ground distance detection method is applied to a depth-controlled planter which is designed for conservation agriculture, the test environment is shown in Fig. 12 (a). Fig. 12 (b) presents the actual distance from the camera to the ground, measured with the measuring tape, which is approximately 585.0mm. Fig. 12 (c) shows RGB image in the ROI acquired by the color camera. Fig. 12 (d) shows original depth image in the ROI, which is presented in the form 3D point cloud by post processing tool. Fig. 12 (e) shows the depth image after removing the influence of residues, and the ground distance detection result by the proposed method is 583.4mm. Compared with the measured distance (585.0mm), the error is 1.6mm. Fig. 12 (f) shows the depth image after removing residues in the form of point cloud, it can be seen some local areas are not fully removed.

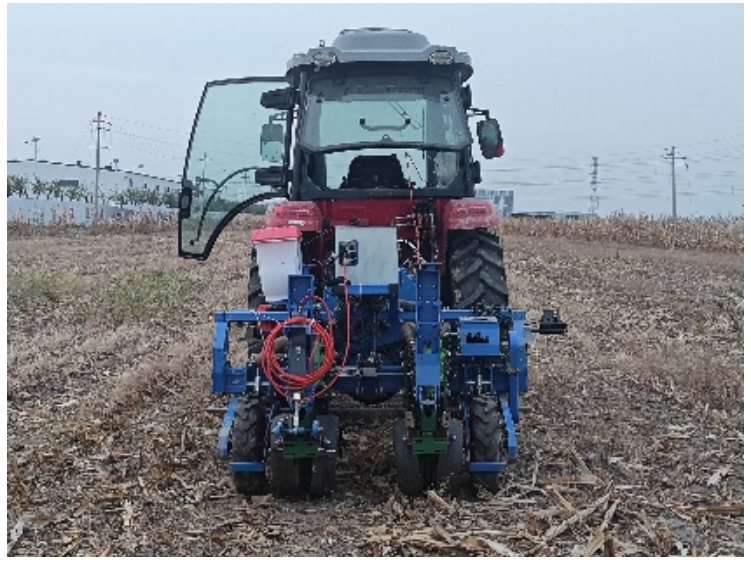

(a) The field test environment

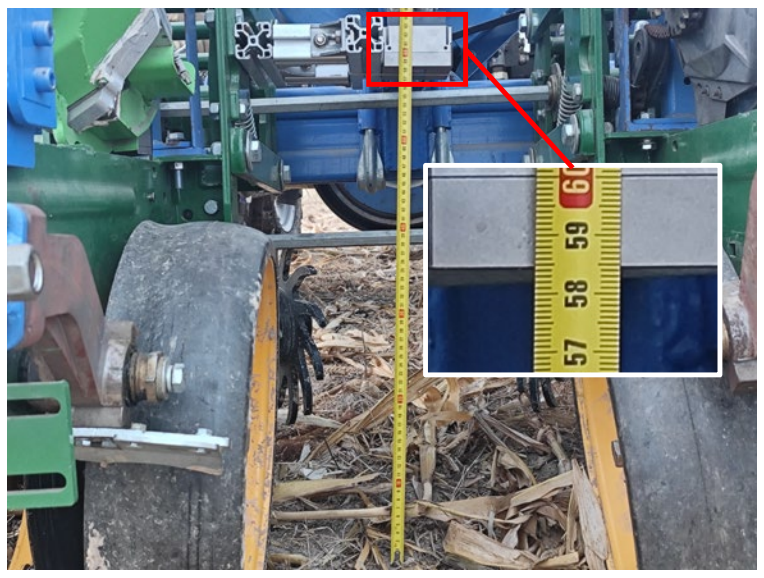

(b) Configuration and actual distance

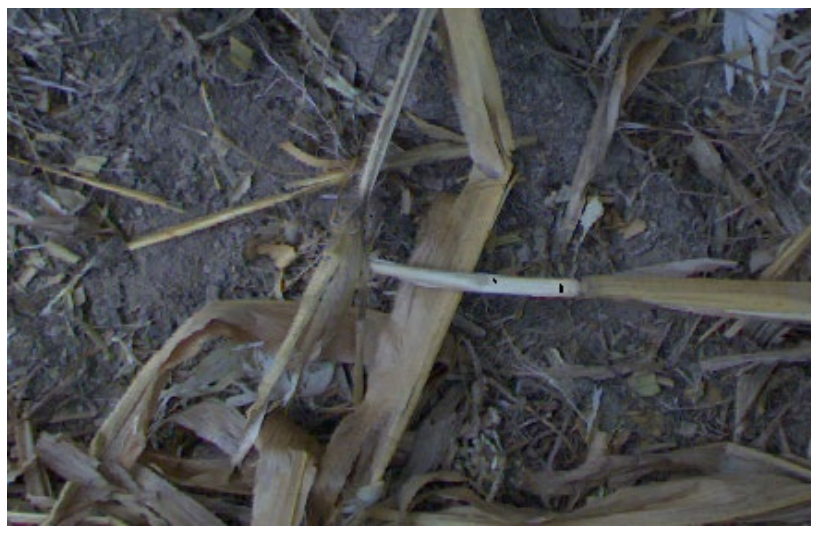

(c) RGB image in ROI

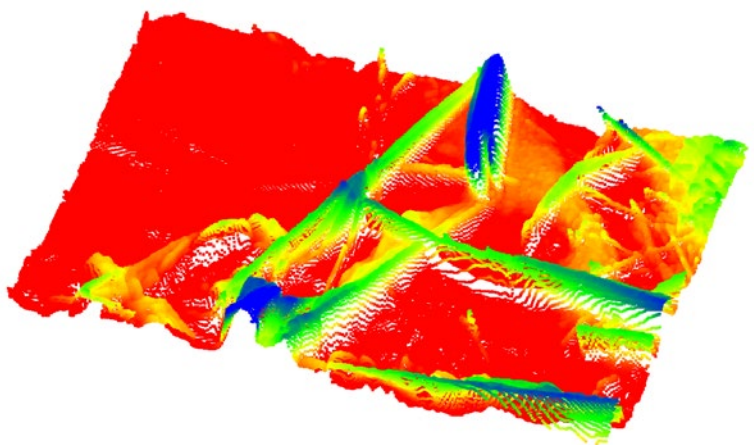

(d) Depth image in ROI shown in the form of point cloud

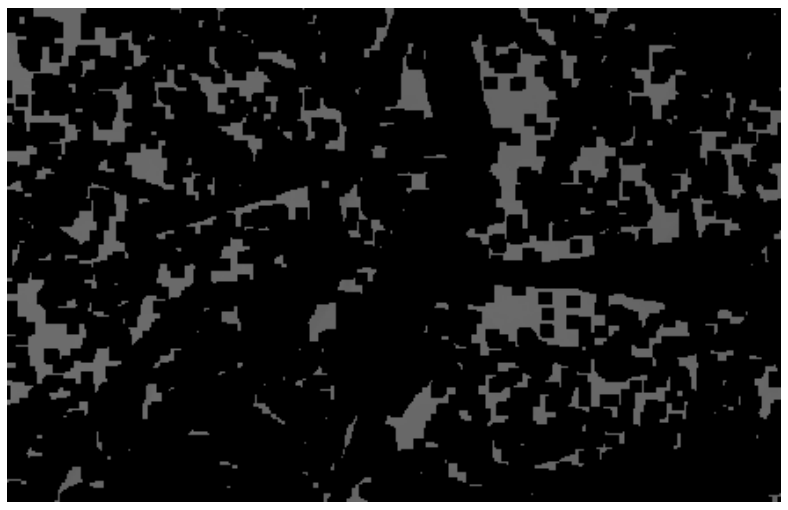

(e) Depth image after removing residues

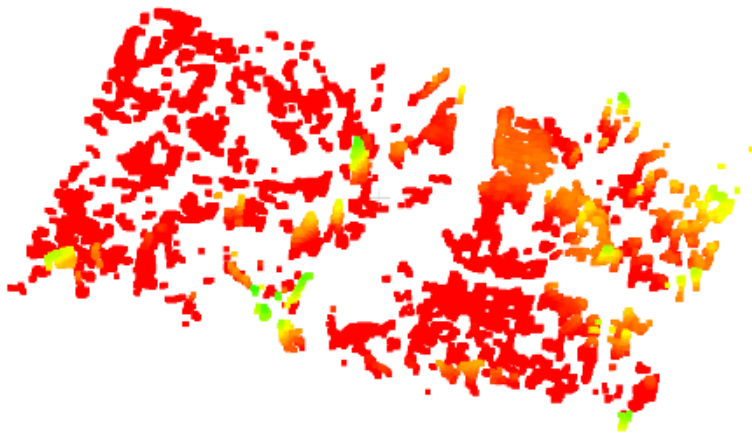

(f) Depth image after removing residues (by point cloud)

Fig. 12 Ground distance detection of the planter using the method in this paper

## Conclusions

(1) A novel Image-based method for detecting the ground distance of soil covered with crop residues has been proposed. It has overcome the limitations of conventional ground distance measurement methods, which lack the ability to distinguish the distance from the soil or the residues.

(2) The fusion of information from 3D and RGB cameras enables effective measurement, eliminating the influence of crop residues and achieving accurate ranging in complex soil surface environments.

(3) The method is experimentally validated under both static, dynamic and field conditions. The results demonstrate that the proposed approach can effectively detect ground distance in complex ground environments, the error cloud be less than 3mm.

(4) One limitation is that when the covering is overly thick to the extent that the ground becomes completely invisible, the ground distance cannot be measured. However, the physical characteristics of the residues along with the movement caused by machinery make such a situation quite rare. Besides, the fully covered image frames can also be ignored.

(5) This method makes it possible to achieve precise depth control in the complex soil conditions of conservation agriculture in the future. It can be applied to various types of machinery, including those used for sowing, tillage, transplanting, and so on. This method has the potential to contribute to the development of precision agriculture and sustainable agriculture.

## Acknowledgements

This research was supported by the National Key R&D Program of China (Grant No. 2022YFE0125800), Shandong Provincial Natural Science Foundation, China (Grant No. ZR2021ME018), as well as Collaborative Innovation Center for Shangdong's Main crop Production Equipment and Mechanization (SDXTZX-22). Any views, findings, conclusions or recommendations expressed in this publication are those of the author.